\documentclass[conference]{IEEEtran}
\IEEEoverridecommandlockouts
\usepackage{ulem}
\usepackage{booktabs}
\usepackage{amsmath,amssymb,amsfonts}
\usepackage{algorithmic}
\usepackage{textcomp}
\usepackage{xcolor}
\usepackage{yfonts}
\usepackage[pdftex]{graphicx}
\graphicspath{{./Figures/}}
\usepackage[backend=biber,style=ieee]{biblatex}
\begin{document}
	\title{Parameter Reduction in Probabilistic Critical Time Evaluation Using Sensitivity Analysis and PCA\\
		\thanks{This study was financed in part by the Coordena\c{c}\~ao de Aperfei\c{c}oamento de Pessoal de N\'ivel Superior -- Brasil (CAPES) -- Finance Code 001.}}

	%% To specify the authors when (number of affiliations <= 2)
	\author{
		\IEEEauthorblockN{1\textsuperscript{st} Raphael Luiz Vicente Fortulan}
		\IEEEauthorblockA{Materials and Engineering Research Institute\\
			Sheffield Hallam University\\
			Sheffield, UK\\
			raphael.l.fortulan@student.shu.ac.uk
	}
	\and
	\IEEEauthorblockN{2\textsuperscript{nd}  Lu\'is Fernando Costa Alberto}
		\IEEEauthorblockA{Department of Electrical and Computer Engineering\\
	University of S\~ao Paulo\\
	S\~ao Carlos, Brazil\\
	lfcalberto@usp.br
					}
				}
	
	\maketitle
	
	\begin{abstract}
		In this paper, we discuss a method to find the most influential power system parameters to the probabilistic transient stability assessment problem---finding the probability distribution of the critical clearing time. We perform the parameter selection by employing a sensitivity analysis combined with a principal component analysis. First, we determine the sensitivity of the machine angles with respect to all system parameters. Second, we employ the principal component analysis algorithm to identify the most influential parameters in the transient stability problem. By identifying such parameters, we can reduce the number of uncertain parameters to only the influential ones in the probabilistic assessment of transient stability, providing a significant speedup in the probabilistic analysis of large power systems. \\
		The proposed algorithm was tested in the IEEE 14 bus systems and the results obtained show that our method can effectively find the most influential parameters.
	\end{abstract}
	\begin{IEEEkeywords}
		probabilistic transient stability analysis; principal component analysis; sensitivity analysis.
	\end{IEEEkeywords}

	\section{Introduction}
    Stability analysis constitutes a necessary step to guarantee the continuous and reliable operation of electric power systems. A power system is safe if it can withstand a predefined list of contingencies, without violating operational limits \cite{Kundur2004}.\\
	With crescent adoption of intermittent and variable generation, such as solar and wind, the use of deterministic methods to analyse the transient stability of power systems fails to capture the system profile. Therefore, the critical clearing time---the largest time interval for the protection activation to ensure stability of the post-fault system---given by a deterministic assessment is inaccurate.\\
	A simplistic solution to the problem is to employ the Monte Carlo method to find the probability distribution of the critical clearing time, considering a set of uncertain parameters. The uncertain parameters can be, for example, the transmission line resistance and reactance. In this work, the primary source of uncertainty of the power system will be the loads, since the variable renewable generation will influence their value.\\
	A glaring problem of the Monte Carlo method is its execution time, since the method samples the random variables---in this case, the uncertain parameters; performs a deterministic analysis for each sample; and, finally, aggregates the obtained results. Two mains factors contribute to slow down the analysis via Monte Carlo Method: the variance of the random variables and the number of random variables. While the former is intrinsic to the problem, the latter can be addressed, and it is the focus of discussion of this paper.\\
	To achieve the reduction of random variables in the Monte Carlo method, we propose the use of the sensitivity of the machine angles---as they have greater influence over the critical clearing time---with respect to the system parameters and the principal component analysis~\cite{Bard1974} to find the most influential parameters to the probabilistic transient stability problem. Considering only the most influential parameters in the probabilistic assessment speeds up the analysis and offer a better understanding of the power system behaviour.\\
	The proposed sensitivity analysis was tested in the IEEE 14 bus system. The test was conducted according to the following sequence: First, the most influential parameters for transient stability analysis were selected using our proposed methodology. Afterwards, the Monte Carlo method was employed to assess the transient stability of the power system in two situations: 1) with every parameter being uncertain; and 2) with only the most influential parameters being uncertain.  The obtained results were compared and it was verified that the proposed methodology gives an excellent result, successfully identifying the most influential parameters in transient stability.\\
	This paper shows a novel use of both sensitivity analysis and principal component analysis since, in the literature, both methods are treated separately. For example, in~\cite{king2016independent}, both the independent component and principal component analysis are used in a machine-learning algorithm to estimate the critical clearing time. While, in~\cite{sharma2018sensitivity}, the sensitivity of the critical clearing time with respect to any system parameter is used to avoid recomputing the critical clearing time for each parameter change. \\
	\hfill\\
	This paper is divided as follows:
	\begin{itemize}
		\item In section~\ref{sec:Transient}, we briefly review the transient stability problem;
		\item In section~\ref{sec:Sensitivity}, we review the sensitivity analysis of a general system;
		\item In section~\ref{sec:PCA}, we introduce the principal component analysis; 
		\item In section~\ref{sec:Examples}, three cases are studied to illustrate the proposed methodology;
		\item In section~\ref{sec:Conclusion} conclusions are drawn.
	\end{itemize}
	\section{Transient Stability}\label{sec:Transient}
	In this section, we review the transient stability problem. A power system experiencing a disturbance can be modelled by a set of three differential equations \cite{kundur1994power, Fortulan}:
	\begin{align}
		&x'_{prf}(t)=f^{prf}\left(x_{prf}(t), \lambda_{prf} \right) \quad{} t\in (-\infty,0],\\
		&x'_{f}(t)=f^{f}\left(x_{f}(t),\lambda_{f} \right) \quad{} t\in (0,t_{cl}],\\
		&x'_{pf}(t)=f^{pf}\left(x_{pf}(t),\lambda_{pf} \right) \quad{} t\in (t_{cl},\infty),
	\end{align}
    where $x_{prf},x_{f},x_{pf}$; $f^{prf},f^{f},f^{pf}$; and $\lambda_{prf},\lambda_{f},\lambda_{pf}$ are,
    respectively, the state variables (angle and frequency); the vector fields; and the parameters of the system at pre-fault, fault-on, and post-fault periods; $t_{cl}$ is the fault clearing time. The analysis is divided into three different periods. At $t=0$, the system experiences a fault and the dynamics change from the one driven by $f^{prf}$ to $f^{f}$. For $t\in (0, t_{cl}]$, the fault-on period, the system dynamics is ruled by $f^{f}$. When protection acts and the fault is cleared, the post-fault period initiates. During this period the system is governed by $f^{pf}$.\\
	The main objective of the transient stability assessment is to find the smallest value for $t_{cl}$, such as the post-fault system remains stable. This value is known as the critical clearing time or $t_{cr}$. In particular, in a probabilistic transient stability assessment, we want to find the probability distribution of $t_{cr}$ given that our pre-fault parameters are random variables.
	\section{Sensitivity Analysis}\label{sec:Sensitivity}
	In this section, we review the sensitivity analysis problem, as it is the first step to find the most influential power system parameters to the transient stability assessment.\\
	In general, for transient stability analysis a power system is modelled by the following initial value problem:
	\begin{equation}
		x'
 = f(x,\lambda),\quad x(t_0)=x_0,
		\label{eq:Sense1}
	\end{equation}
	where $x\in\mathbb{R}^n$ and $\lambda\in\mathbb{R}^m$ are, respectively, the system state and parameter vectors. The value of $x$ at $t$ is represented by:
	\begin{equation}
		x(t,\lambda).
	\end{equation}
	The effect of the parameters in~\eqref{eq:Sense1} can be investigated by adding $\Delta\lambda_{k}$ to the $k$-th parameter $\lambda_{k}$ \cite{Frank1980, tomovic1963sensitivity}. Since $\lambda$ is a vector, we adopt the following notation:
	\begin{equation}
		\textswab{p}_k = \begin{bmatrix}
			0\\
			\vdots\\
			\Delta\lambda_{k}\\
			\vdots\\
			0
		\end{bmatrix},
	\end{equation}
	where $\textswab{p}_k$ is a perturbation vector. The Taylor series of the perturbed solution is as follows:
	\begin{equation}
		x(t,\lambda+\textswab{p}_k) = \sum_{j=0}^{\infty} \left[ \frac{1}{j!}\left( \Delta\lambda_{i} \frac{\partial}{\partial \lambda_{i}}\right)^{j} x(t,\lambda)\right].
		\label{eq:Sense2}
	\end{equation}
	Truncating~\eqref{eq:Sense2} in the first term yields:
	\begin{equation}
		x(t,\lambda+\textswab{p}_{k}) - x(t,\lambda) \approx \frac{\partial x(t,\lambda)}{\partial\lambda_{k}}\Delta\lambda_k\rightarrow \Delta x\approx s_k\Delta\lambda_k,
		\label{eq:Sense3}
	\end{equation}
	where $s_k$ is called the local sensitivity of $x$ with respect to the parameter $\lambda_k$. Naturally, the local sensitivity of $x_i$ with respect to the parameter $\lambda_k$ is:
	\begin{equation}
		s_{ik} = \frac{\partial x_{i}(t,\lambda)}{\partial \lambda_k}.
		\label{eq:Sense4}
	\end{equation}
	The sensitivity matrix is then given by:
	\begin{equation}
		S=(s_{ik}),
		\label{eq:Sens}
	\end{equation}
	where $i=1,2,\hdots, n$ and $k = 1,2,\hdots, m$.\\
	The normalised sensitivity is defined as follows:
	\begin{equation}
		\tilde{s}_{ik} = \frac{\lambda_{k}}{x_{i}}\frac{\partial x_{i}(t,\lambda)}{\partial \lambda_k} = \frac{\partial\ln(x_{i}(t,\lambda))}{\partial\ln(\lambda_k)}.
	\end{equation}
	Likewise, the normalised sensitivity matrix is given by:
	\begin{equation}
		\tilde{S}=(\tilde{s}_{ik}),
		\label{eq:NormSens}
	\end{equation}
	where $i=1,2,\hdots, n$ and $k = 1,2,\hdots, m$.\\
	\subsection{Sensitivity Analysis in Transient Stability}
	In the transient stability problem, the parameter vector $\lambda$ will be composed by the system loads, line parameters and generator parameters. The system states will be the generator angle and frequency.\\
	To simplify our analysis we considered that only the line parameters and loads have uncertainties, as the machine parameters are generally well known. We also only evaluated the sensitivity of the generator angle, as it dominates over the $t_{cr}$ value. Thus, for a general power system with $m$ loads, $l$ lines, and $n$ machines:
	\begin{align}
		\delta &= [\delta_{1},\delta_{2},\hdots,\delta_{n}]^{\mathsf{T}}\rightarrow\text{Machine angles},\\
		P_L &= [P_{L_1},P_{L_2},\hdots,P_{L_m}]^{\mathsf{T}}\rightarrow\text{Active loads},\\
		Q_L &= [Q_{L_1},Q_{L_2},\hdots,Q_{L_m}]^{\mathsf{T}}\rightarrow\text{Reactive loads},\\
		R &= [R_{1},R_{2},\hdots,R_{l}]^{\mathsf{T}}\rightarrow\text{Line resistances},\\
		X &= [X_{1},X_{2},\hdots,X_{l}]^{\mathsf{T}}\rightarrow\text{Line reactances},\\
		B/2 &= [B_{1},B_{2},\hdots,B_{l}]^{\mathsf{T}}\rightarrow\text{Shunts},
	\end{align}
	with the sensitivity matrix and normalized sensitivity matrix being evaluated by~\eqref{eq:Sens} and~\eqref{eq:NormSens} respectively.\\
	Finally, because the critical clearing time is primarily controlled by the fault-on system behavior, we evaluated the sensitivity/normalized sensitivity matrix over the fault-on trajectory.
	\section{PCA - Principal Component Analysis}\label{sec:PCA}
To quantify the effect of several parameters changing at the same time, consider the following loss function~\cite{vajda1985principal}:
	\begin{equation}
		\ell (p) = \int_{0}^{t} \sum_{i=1}^{n} \left(\frac{x_i (p,\tau)-x_i(p_{0},\tau)}{x_i(p_{0},\tau)}\right)^2 \mathrm{d}\tau,
        \label{eq:jp}
	\end{equation}
	where \(x_i(p,t)\) and $x_i(p_{0},t)$ are, respectively, the perturbed and nominal solutions;  $p=\ln(\lambda)$; and $\ell(p)$ measures the normalized error between the perturbed and nominal solutions over time.\\
    Assuming that both solutions are sampled in \textswab{r} points, we obtain:
	\begin{multline}
		\ell (p) = \int_{0}^{t} \sum_{i=1}^{n} \sum_{j=0}^{\textswab{r} - 1}\left(\frac{x_i (p,\tau)-x_i(p_{0},\tau)}{x_i(p_{0},\tau)}\right)^2 \delta(\tau-j)\mathrm{d}\tau =\\ \sum_{j=0}^{\textswab{r} - 1} \sum_{i=1}^{n}\left[\frac{x_i (p,t_j)-x_i(p_{0},t_j)}{x_i(p_{0},t_j)}\right]^2,
	\end{multline}
	where \(\delta(\cdot)\) is the Dirac delta function. The number of sampled points will be determined by the numerical method used to solve the differential equations of the fault-on power system.\\
	The Taylor series for $\ell (p)$ centred at $p_{0}$ is as follows:
	\begin{equation}
		  \ell (p) = \ell (p_{0})+ \nabla  \ell (p_{0}) (p-p_{0}) +\frac{1}{2}(p-p_{0})^\mathsf{T} \nabla^2 \ell (p_{0})(p-p_{0})+\cdots.
	\end{equation}
	Since $\ell (p_{0})=0$ and it is a minimum, which implies in $\nabla \ell (p_{0}) = 0$:
	\begin{equation}
		\ell (p) \approx \frac{1}{2}(p-p_{0})^\mathsf{T} \nabla^2 \ell (p_{0})(p-p_{0}).
		\label{eq:PCA1}
	\end{equation}
	The Hessian matrix $\nabla^2l(p_))$ expression is obtained as follows:
	\begin{multline}
		\ell (p) = \sum_{j=0}^{\textswab{r} - 1} \sum_{i=1}^{n}\left[\frac{r_i (t_j)}{x_i(p_{0},t_j)}\right]^2 \rightarrow\\ \nabla \ell (p)_k = \sum_{j=0}^{\textswab{r} - 1} \sum_{i=1}^{n} \frac{\partial}{\partial p_k}\left(\left[\frac{r_i (t_j)}{x_i(p_{0},t_j)}\right]^2\right),
		\label{eq:PCA2}
	\end{multline}
	where $r_{i}(t_j)$ is the $i$-th residue at a time $t_j$. Applying the chain rule in~\eqref{eq:PCA2} and neglecting the second-order derivative terms:
	\begin{multline}
		\nabla^2 \ell (p)_{ks} \approx 2\sum_{j=0}^{\textswab{r} - 1} \sum_{i=1}^{n} \frac{\partial}{\partial p_s}\left(\frac{r_i (t_j)}{x_i(p_{0},t_j)}\right)\frac{\partial}{\partial p_k}\left(\frac{r_i (t_j)}{x_i(p_{0},t_j)}\right).
		\label{eq:PCA4}
	\end{multline}
	Since $r_{i}(t_j) = x_i (p,t_j)-x_i(p_{0},t_j)$:
	\begin{multline}
		\frac{\partial}{\partial p_k}\left(\frac{r_i (t_j)}{x_i(p_{0},t_j)}\right) = \frac{\partial}{\partial p_k}\left(\frac{x_i (p,t_j)}{x_i(p_{0},t_j)}-1\right) =\\ \frac{1}{x_i(p_{0},t_j)} \frac{\partial x_i (p,t_j)}{\partial p_k} \xrightarrow[p=p_0]{\text{at}} \frac{\partial \ln(x_i(p_0,t_j))}{\partial p_k}.
	\end{multline}
	\\ \hfill
	\\ \hfill
	Thus,
	\begin{equation}
		\nabla^2 \ell(p_0) = 2\sum_{j=0}^{\textswab{r} - 1} \tilde{S}_{j}^{\mathsf{T}} \tilde{S}_{j}\rightarrow \ell(p) \approx (p-p_{0})^\mathsf{T} \sum_{j=0}^{\textswab{r} - 1} \tilde{S}_{j}^{\mathsf{T}} \tilde{S}_{j}(p-p_{0}),
	\end{equation}
	where $\tilde{S}_{j}$ is the normalised sensitivity matrix at $t_j$.\newline
	To understand how $\ell(p)$ varies with $p$, the eigendecomposition of $\sum_{j=0}^{\textswab{r} - 1} \tilde{S}_{j}^{\mathsf{T}} \tilde{S}_{j}$ was employed  as follows~\cite{Bard1974}:
	\begin{equation}
		\sum_{j=0}^{\textswab{r} - 1} \tilde{S}_{j}^{\mathsf{T}} \tilde{S}_{j} = U\Pi U^\mathsf{T},
	\end{equation}
	where $\Pi$ is the diagonal matrix of the eigenvalues $\pi_i$ of $\sum_{j=0}^{\textswab{r} - 1} \tilde{S}_{j}^{\mathsf{T}} \tilde{S}_{j}$ and $U$ is the square matrix whose $i$-th column is the $u_i$ normalized eigenvector of $\sum_{j=0}^{\textswab{r} - 1} \tilde{S}_{j}^{\mathsf{T}} \tilde{S}_{j}$. Therefore, $\ell(p)$ is most sensitive to changes in $p$ along the eigenvector corresponding to the largest eigenvalue. As such, literature refers to $\rho = U^{\mathsf{T}}(p-p_0)$ as the principal components.\newline
	The most influential parameters can be identified as follows: let $\rho_{\max}$ be the principal component corresponding to the largest eigenvalue, $\pi_{\max}$, of $\sum_{j=0}^{\textswab{r} - 1} \tilde{S}_{j}^{\mathsf{T}} \tilde{S}_{j}$, then:
	\begin{equation}
		\rho_{\max} = u_{\max}^{\mathsf{T}} (p-p_0),
	\end{equation}
	where $u_{\max}$ is the eigenvector corresponding to $\pi_{\max}$. Since $u_{\max}$ is unitary, the percentage influence of the variation of each parameter can be estimated.\\
	Clearly, for a large system it is unpractical to calculate all eigenvalues. Luckily, as only the largest eigenvalue and its correspondent eigenvector are of our interest, fast methods such as Power Iteration \cite{Burden2010} algorithm can be employed to find them.
	\section{Examples}\label{sec:Examples}
\begin{figure}[!ht]
		\centering
		\includegraphics[scale=0.05]{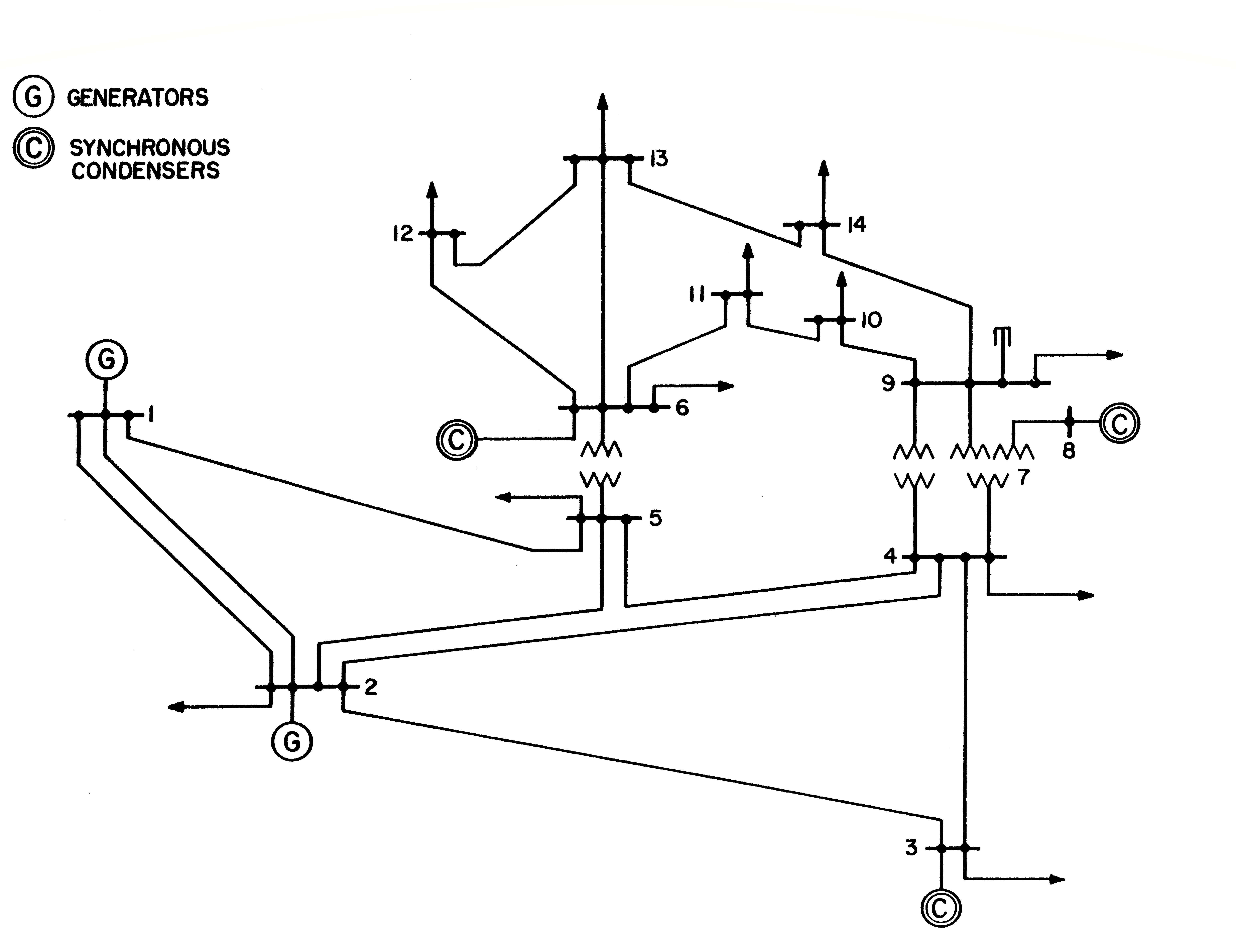}
		\caption{IEEE 14 bus system}
		\label{fig:14bus}
	\end{figure}

	In this section, several examples showing that our methodology can really identify the most influential parameters to the probabilistic transient stability assessment are presented.\\
	In all examples, we considered that the parameters follow a normal distribution with mean equal to their nominal value and a coefficient of variation of 5\% to all loads and 2.5\% to all lines parameters. The coefficient of variation can be defined as the ratio of the standard deviation $\sigma$ to the mean $\mu$~\cite{Cambridge}:
	\begin{equation}
		c_v = \frac{\sigma}{\mu}.
	\end{equation}
	All examples were executed in the IEEE 14 bus system \cite{14busWashington}, depicted in Figure~\ref{fig:14bus}.
	\subsection{Case I}
	In this case, a solid three-phase short-circuit occurs at bus 1 and the fault is cleared by opening line 1-5. Using our methodology, we identified that the following parameters have 97.5\% influence over the critical clearing time: $R_{1-2},
 P_{L_3},  X_{1-2}, P_{L_4}, P_{L_9},P_{L_5},P_{L_2},B_{1-2},X_{1-5}$, where $R_{i-j}$ and $X_{i-j}$ represents, respectively, the resistance and reactance between the buses $i$ and $j$; and $P_{L_i}$ represents an active load located at bus $i$.\\
	To verify the accuracy of our analysis, we first evaluate the critical clearing time applying uncertainties in all parameters and then only in the most influential ones. Both evaluations were executed using a Monte Carlo simulation. 
	\begin{table}[!ht]
		\centering
		\caption{Critical Clearing Time - Case I}
		\begin{tabular}{@{}lll@{}}
			\toprule
			Uncertainties                & $\mu$    & $\sigma$ \\ \midrule
			In all parameters            & 0.3617   & 0.009815 \\
			Only in the most influential & 0.3670   & 0.009047 \\ \bottomrule
		\end{tabular}
		\label{tab:CaseI}
	\end{table}\\
	Table~\ref{tab:CaseI} shows the simulation with only the most influential parameters being uncertain contains 92.2\% of the total variance---obtained in the simulation with uncertainties in all parameters. Thus, our methodology successfully identified the most influential parameters.\\
	Clearly the simulation result differs from the expected 97.5\%, this, however, was anticipated for two reasons. First, we are using a linear analysis in a highly non-linear problem. Second, we are not evaluating the sensitivity of the critical clearing time, but the sensitivity of the generator angles with respect to the parameters.
	\subsection{Case II}
	In this case, a solid three-phase short-circuit occurs at bus 9 and the fault is cleared by opening line 4-9. Using our methodology we identified that the following parameters have 95\% influence over the critical clearing time:$P_{L_3},P_{L_4}, R_{1-5}, P_{L_9} ,X_{5-6} P_{L_5}, P_{L_2} ,X_{1-2} ,X_{6-11},$ $X_{10-11},X_{4-9},X_{13-14} ,X_{4-7} ,R_{2-5}, R_{6-11}, X_{2-4}, R_{2-4},$ $X_{1-5}, R_{10-11}, X_{9-14}$.\\
	To verify the accuracy of our analysis, we proceed as in Case I. 
	\begin{table}[!ht]
		\centering
		\caption{Critical Clearing Time - Case II}
		\begin{tabular}{@{}lll@{}}
			\toprule
			Uncertainties                & $\mu$    & $\sigma$ \\ \midrule
			In all parameters            & 0.5958   & 0.012862 \\
			Only in the most influential & 0.5961   & 0.012529 \\ \bottomrule
		\end{tabular}
		\label{tab:CaseII}
	\end{table}\\
	Table~\ref{tab:CaseII} shows the simulation with only the most influential parameters being uncertain contains 97.4\% of the total variance---obtained in the simulation with uncertainties in all parameters. Thus, our methodology successfully identified the most influential parameters.
	\subsection{Case III}
	In this case, a solid three-phase short-circuit occurs at bus 2 and the fault is cleared by opening line 2-5. Using our methodology we identified that the following parameters have 99\% influence over the critical clearing time: $X_{3-4}, X_{2-3}, R_{2-3}, R_{3-4}, P_{L_3}$.\\
	To verify the accuracy of our analysis, we proceed as in Case I. 
	\begin{table}[!ht]
		\centering
		\caption{Critical Clearing Time - Case III}
		\begin{tabular}{@{}lll@{}}
			\toprule
			Uncertainties                & $\mu$    & $\sigma$ \\ \midrule
			In all parameters            & 1.4058   & 0.273488 \\
			Only in the most influential & 1.4040   & 0.272346 \\ \bottomrule
		\end{tabular}
		\label{tab:CaseIII}
	\end{table}\\
	 Table~\ref{tab:CaseIII} shows the simulation with only the most influential parameters being uncertain contains 99.58\% of the total variance---obtained in the simulation with uncertainties in all parameters. Thus, our methodology successfully identified the most influential parameters.\\
	\section{Conclusions}\label{sec:Conclusion}
	In this paper, we introduced a methodology to find the most influential parameters for the probabilistic transient stability problem. In particular, we apply the widely used machine learning method called principal component analysis combined with a sensitivity analysis of the generator angles to find the most influential parameters.\\
	The obtained results showed that although we are applying a linear analysis in a highly non-linear system and not evaluating the sensitivity of the critical clearing time per se, we were able to estimate the overall influence of a set of parameters in the transient stability assessment and, by consequence, find the most influential parameters.\\
	Future research will be focused on expanding the methodology to new applications such as voltage stability analysis and control.
	\printbibliography
\end{document}